\documentclass[cernpreprint, atlasdraft=false, UKenglish, texmf, orcidlogo]{atlasdoc}
\usepackage{atlaspackage}
\usepackage{atlasbiblatex}

\usepackage{atlasphysics}
\usepackage{atlasjetetmiss}
\graphicspath{{logos/}{figures/}}

\usepackage{ANA-EXOT-2022-32-PAPER-defs}
\usepackage{graphicx}
\usepackage{multirow}
\AtlasTitle{Search for electroweak scale dijet resonances in pile-up collisions at $\sqrt{s}=13$ TeV with the ATLAS detector}

\AtlasAbstract{%
A search for dijet resonances in the mass range of 100--250 GeV is presented using proton--proton collision data recorded by the ATLAS experiment at the Large Hadron Collider at a centre-of-mass energy of 13 TeV. Conventional searches for hadronic resonances in the sub-TeV regime are heavily constrained by high jet trigger thresholds required to manage the overwhelming rate of Standard Model multijet background processes described by quantum chromodynamics. To overcome this limitation, this analysis uses a novel strategy that separately reconstructs multiple proton--proton interactions per bunch crossing, known as pile-up collisions. The dataset was collected by the ATLAS experiment using single-electron and single-muon triggers in 2016--2018 corresponding to an integrated luminosity of 1.30 pb$^{-1}$, which represents the effective luminosity of pile-up collisions recorded alongside triggered events. This constitutes the first application of pile-up collisions to probe low dijet mass scales. No significant excess is observed above the background expectation and exclusion limits are set for a generic Gaussian model and a simplified dark matter model featuring an axial-vector mediator with coupling to quarks.
}

\AtlasRefCode{EXOT-2022-32}

\PreprintIdNumber{CERN-EP-2026-147}

\AtlasJournal{EPJ C}
\AtlasCoverEgroupAnalysisTeam{atlas-exot-2022-32-analysis-team@cern.ch}

\hypersetup{pdftitle={ATLAS document},pdfauthor={The ATLAS Collaboration}}

\begin{document}

\maketitle

\tableofcontents

\section{Introduction}
\label{sec:intro}

% The next lines are included from the .//sections/Introduction.tex input file
Many extensions to the Standard Model (SM) predict the existence of new particles that couple to quarks and gluons, including excited quarks~\cite{Baur:1987ga}, leptophobic $Z'$ and coloron resonances~\cite{Dobrescu:2013cmh,Chala:2015ama}, and axion-like particles~\cite{Bauer:2017ris}. The hadronic decays of such particles produced in proton--proton (\ppp) collisions would manifest as an excess of events in the invariant mass spectrum of pairs of jets above a smoothly falling background dominated by multijet processes described by quantum chromodynamics (QCD). Consequently, the search for dijet resonances has been a key part of the Large Hadron Collider (LHC)~\cite{Evans:2008zzb} physics programme.
Historically, searches at the LHC targeted heavy resonances in the multi-TeV regime~\cite{EXOT-2019-03, EXOT-2013-11, EXOT-2010-07, CMS-EXO-19-012, CMS-EXO-11-015, CMS-EXO-10-010}. In this mass range, the decay particles from the posited resonance are highly energetic and straightforward to select using standard jet triggers.

Probing resonances below the TeV scale, however, presents significant experimental challenges. In this regime, the production rate of background SM QCD dijet events becomes prohibitive, significantly exceeding the storage bandwidth available to the experiments. To circumvent this limitation, recent analyses by the ATLAS~\cite{PERF-2007-01} and CMS~\cite{CMS-CMS-00-001} collaborations have employed alternative strategies. One such approach involves storing only trigger-reconstructed objects with reduced information for subsequent offline analysis, in order to lower the data transfer rate~\cite{EXOT-2018-39, CMS-EXO-23-004}. Alternatively, searches target the associated production of resonances with a high transverse momentum (\pt) initial-state radiation (ISR) object corresponding to a jet or photon which is used to trigger on the collision~\cite{EXOT-2019-01, EXOT-2022-33, EXOT-2018-05, CMS-EXO-17-027}. While highly effective, these approaches have their intrinsic limitations: storing trigger objects means an analysis cannot rely on offline event reconstruction, whereas ISR-based searches suffer from reduced acceptance and require specific topological assumptions about the signal.

In this paper, a novel search strategy is presented that exploits a unique dataset whereby multiple \ppp{} collisions (\pileup) are reconstructed separately in each bunch crossing recorded by ATLAS during the Run~2 data-taking period~\cite{JETM-2023-08}; this dataset has been demonstrated to be trigger-unbiased. By reconstructing dijet events from these \pileup collisions, which are recorded simultaneously alongside triggered events, the low-mass regime can be accessed while retaining full offline reconstruction capabilities. This enables, for the first time, a search for dijet resonances in \pileup{} collisions.
\Pileup{} collisions are available only for the subset of LHC events \textit{selected} by the trigger system, therefore the effective integrated luminosity is reduced compared to the total luminosity \textit{processed} by the ATLAS trigger system and considered for a traditional trigger-based analysis.
Despite the reduced luminosity, the dataset offers competitive reach for signatures with a high production cross section and a low (jet) trigger acceptance, such as low-momentum multijet events, for which the trigger prescales are often very high~\cite{TRIG-2016-01}.
Other strategies to collect trigger-unbiased data can produce datasets with the same physical properties but with much reduced statistical power.
For example, the pile-up dataset has a factor of 30 higher luminosity than data collected in the ATLAS ZeroBias stream~\cite{TRIG-2016-01}.

The results of the analysis are interpreted in the context of SM $W/Z$ production, a simplified $s$-channel dark matter model involving a leptophobic axial-vector mediator, $Z^\prime$, and a generic Gaussian signal in the mass range 100--250 GeV.
A model is chosen in which the $Z^\prime$ is allowed to decay only into a quark pair, as the dark matter particle mass is assumed to lie significantly above the kinematically accessible phase space.
No assumptions are made for the final states of the Gaussian resonances, which can also include gluon jets.

The probed invariant mass region is already covered by previous searches targeting low-mass resonances, namely the dijet ISR photon~\cite{EXOT-2022-33} search. This earlier search relied on a boosted topology, where the dijet resonance is reconstructed as a single large-radius jet and selections on jet substructure are used to identify \pt-balanced two-prong decays in order to suppress the QCD background. These substructure selections make such searches less sensitive to models in which the decay is not balanced, for example in dark QCD scenarios where part of the shower is invisible~\cite{Cohen:2015toa}. In contrast, the resolved dijet strategy used in this analysis does not rely on substructure selections and therefore retains sensitivity to such signatures. Sensitivity to dark QCD depends on model parameter choices: some parameter combinations result in wide resonances that are similar to the larger width options of the $Z^\prime$ resonances searched for in this result. In addition, this search has unique sensitivity to resonances that couple only to gluons: gluon--gluon initial states produce negligible photon ISR at leading order because gluons are electrically neutral, reducing strongly the sensitivity of the dijet ISR photon search to gluon-initiated production.

The paper is organized as follows. Section \ref{sec:detector} briefly describes the ATLAS detector. The \pileup dataset and simulated samples are described in Section \ref{sec:DataMC}. The reconstruction of physics objects, including \pileup-specific jet reconstruction, is detailed in Section \ref{sec:ObjReco}. Section \ref{sec:EvSel} outlines the analysis selections used to obtain a trigger-unbiased dataset of \pileup collisions. The background estimation strategy and the treatment of systematic uncertainties are discussed in Section \ref{sec:Methodology} and Section \ref{sec:Uncertainties}, respectively. Finally, the results of the search and its conclusions are presented in Section \ref{sec:result} and Section \ref{sec:conclusion}.

% End of text imported from the .//sections/Introduction.tex input file

%

%
\section{ATLAS detector}
\label{sec:detector}

% The next lines are included from the .//sections/Detector.tex input file
The ATLAS detector~\cite{PERF-2007-01,GENR-2019-02} at the LHC covers nearly the entire solid angle around the collision point.\footnote{\AtlasCoordFootnote}
It consists of an inner tracking detector surrounded by a thin superconducting solenoid, electromagnetic and hadronic calorimeters,
and a muon spectrometer incorporating three large superconducting air-core toroidal magnets.

The inner-detector system (ID) is immersed in a \qty{2}{\tesla} axial magnetic field
and provides charged-particle tracking in the range \(|\eta| < 2.5\).
The high-granularity silicon pixel detector covers the vertex region and typically provides four measurements per track,
the first hit generally being in the insertable B-layer (IBL).
It is followed by the semiconductor tracker (SCT), which usually provides eight measurements per track.
These silicon detectors are complemented by the transition radiation tracker (TRT),
which enables radially extended track reconstruction up to \(|\eta| = 2.0\).
The TRT also provides electron identification information
based on the fraction of hits (typically 30 in total) above a higher energy-deposit threshold corresponding to transition radiation.

The calorimeter system covers the pseudorapidity range \(|\eta| < 4.9\).
Within the region \(|\eta|< 3.2\), electromagnetic calorimetry is provided by barrel and
endcap high-granularity lead/liquid-argon (LAr) calorimeters,
with an additional thin LAr presampler covering \(|\eta| < 1.8\)
to correct for energy loss in material upstream of the calorimeters.
Hadronic calorimetry is provided by the steel/scintillating-tile calorimeter (Tile Calorimeter),
segmented into three barrel structures within \(|\eta| < 1.7\), and two copper/LAr hadronic endcap calorimeters.
The solid angle coverage is completed with forward copper/LAr and tungsten/LAr calorimeter modules
covering $3.2 < |\eta| < 4.9$, optimised for electromagnetic and hadronic energy measurements, respectively.

The muon spectrometer (MS) comprises separate trigger and
high-precision tracking chambers measuring the deflection of muons in a magnetic field generated by the superconducting air-core toroidal magnets.
The field integral of the toroids ranges between \num{2.0} and \qty{6.0}{\tesla\metre}
across most of the detector.
Three layers of precision chambers, each consisting of layers of monitored drift tubes, cover the region \(|\eta| < 2.7\).
These are complemented in the innermost layer of the endcap region by cathode-strip chambers, providing precision tracking in the region \(1.3 < |\eta| < 2.7\).
The muon trigger system covers the range \(|\eta| < 2.4\) with resistive-plate chambers in the barrel region, and thin-gap chambers in the endcap regions.

Events were selected by the first-level trigger system implemented in custom hardware,
followed by selections made by algorithms implemented in software in the high-level trigger (HLT)~\cite{TRIG-2016-01,TRIG-2018-05,TRIG-2018-01}.
The first-level trigger accepted events from the \qty{40}{\MHz} nominal (\qty{30}{\MHz} average filled~\cite{JETM-2023-08}) bunch crossing frequency at a rate close to \qty{100}{\kHz},
which the high-level trigger further reduced in order to record complete events to disk at about \qty{1.25}{\kHz}.

A software suite~\cite{SOFT-2022-02} is used in data simulation, in the reconstruction
and analysis of real and simulated data, in detector operations, and in the trigger and data acquisition
systems of the experiment.

% End of text imported from the .//sections/Detector.tex input file

%

%
\section{Data and simulated samples}
\label{sec:DataMC}

% The next lines are included from the .//sections/Datasets.tex input file
The analysis makes use of \ppp{} collision data collected by ATLAS in 2016--2018. The data are selected using single-electron and single-muon triggers~\cite{TRIG-2018-01, TRIG-2018-05}, with a 26 GeV \pt requirement at the HLT. These triggers are used as they correspond to a significant fraction of the total trigger rate and because the single-lepton signature is relatively easy to identify and match to a given collision vertex.

In this paper the full detector readout following an HLT accept signal is referred to as a recorded bunch crossing (\rbc); the term \textit{event} is instead used to refer to the detector signals associated with a physics process generated by a single \ppp{} collision.
Each \rbc{} contains on average 34 independent \ppp{} collisions, of which 60\% can be reconstructed as separate primary vertices~\cite{IDTR-2021-01}.
In the context of dijet resonances, the difference between \ppp{} collisions and primary vertices is reconstruction-efficiency-based not physics-origin-based, thus the two terms will often be used interchangeably, highlighting the conceptual goal of reconstructing every \ppp{} collision.
Reconstructed primary vertices can be further classified as being either a triggering primary vertex (\tpv) or a \pileup{} primary vertex (\ppv): \tpvs are defined as vertices associated with the \ppp{} interaction(s) that produced the electron and/or muon signature(s) that triggered the recording of the bunch crossing, while all other reconstructed primary vertices are labelled as \ppvs{}.
The \pileup{} dataset is obtained by processing each \rbc{}, identifying and removing the \tpv{}(s) from that \rbc, and analysing the remaining \ppvs{}. Primary vertex reconstruction is described in \Section{sec:ObjReco}, while the identification of the \tpv and its removal are described in more detail in \Section{sec:EvSel}.

The \pileup{} reconstruction procedure yields a recorded dataset corresponding to an effective integrated luminosity of 1.30~\ipb. The integrated luminosity of the \pileup dataset is obtained by considering the combined rate of data \textit{recorded} using the single-electron and single-muon triggers as a fraction of the total \textit{delivered} bunch crossing rate, following the procedure described in Ref.~\cite{JETM-2023-08}. This results in the \pileup dataset integrated luminosity being roughly (30\,MHz delivered)/(300\,Hz recorded) = $10^5$ times smaller than the integrated luminosity of the delivered dataset, which is measured by the LUCID--2 detector.

Monte Carlo (MC) samples of signal and background events are used to optimize the analysis selection and to interpret the results. When using simulated event samples, the assumption is that the generated process of interest would constitute an individual \ppp{}  collision vertex in the \pileup dataset after all quality selection procedures detailed in Ref.~\cite{JETM-2023-08}.
These samples are therefore generated using the standard ATLAS simulation pipeline, without extra requirements on the presence of leptons for the trigger selections or TPV removal.

The main benchmark model is a simplified leptophobic $Z^\prime$ signal, generated according to recommendations in Refs.~\cite{ALBERT2019100377,ABERCROMBIE2020100371}. The signal is assumed to be produced in the $s$-channel with axial-vector interactions. The simplified model is described by two parameters: the mediator mass, $m_{Z^\prime}$, and its coupling to the quarks, $g_q$. A grid of signal samples is generated, including the following points: $m_{Z^\prime} = 80, 100, 150, 250\,\GeV{}$ and an equal coupling to all quarks of $g_q \in [0.2, 1.0]$, with a spacing of 0.2.
The intrinsic width of the $Z^\prime$ is accounted for at all times, and is found to be non-negligible for $g_q$ larger than roughly 0.2, leading to wider resonances for higher $g_q$ values.
Couplings with $g_q >1.0$ are not considered as the generated resonance width-to-mass ratio exceeds 50\%.
The signal samples neglect any interference effects between the $Z'$ and the SM $Z$ boson.
The $Z^\prime$ decays to $u, d, s, c, b$ quark flavours. Decays of the $Z^\prime$ into the dark matter candidate are assumed to be negligible, with the dark matter coupling set to $g_\chi = 1$ and the mass fixed at $m_\chi = 10~\text{TeV}$, making such decays kinematically forbidden.

The signal samples are generated using the \MADGRAPH[2.9.16]~\cite{Alwall:2011uj} generator at leading order (LO) in perturbative QCD with the \NNPDF[2.3lo] parton distribution function (PDF) set~\cite{Ball:2012cx}. The events are interfaced with \PYTHIA[8.307]~\cite{Bierlich:2022pfr} for parton showering, hadronization and underlying event modelling, with parameter values set according to the A14 set of tuned parameters ('tune')~\cite{ATL-PHYS-PUB-2014-021}. \EVTGEN[1.7.0
]~\cite{Lange:2001uf} is used to model the decays of heavy flavour quarks.

Simulated QCD multijet samples are employed to obtain the detector resolution binning for the invariant mass fit and for signal-to-background optimisation studies. Multijet samples were generated using \PYTHIA[8.230]~\cite{Sjostrand:2014zea} with leading-order matrix elements for dijet production, which were matched to the parton shower.
The renormalisation and factorisation scales were set to the geometric mean of the squared transverse masses of the two outgoing particles in the matrix element, \(\pTHatPythia = \sqrt{(\pTX[2][1] + m_1^2) (\pTX[2][2] + m_2^2)}\). The \NNPDF[2.3lo] PDF set~\cite{Ball:2012cx} was used in the matrix-element (ME) generation, the parton shower, and the simulation of the multi-parton interactions. The A14 set of tuned parameters was used.

The generated signal and background events are passed through detector simulation~\cite{SOFT-2010-01} based on \GEANT~\cite{Agostinelli:2002hh}. The effect of \pileup on the process of interest was modelled by overlaying the simulated hard-scattering event with
inelastic \(pp\) events generated with \PYTHIA[8.186]~\cite{Sjostrand:2007gs}
using the \NNPDF[2.3lo] PDF set~\cite{Ball:2012cx} and the
A3 tune~\cite{ATL-PHYS-PUB-2016-017}. The simulated pile-up profile was reweighted to match the distribution of the number of interactions per bunch crossing observed in data.
These overlaid collisions are necessary to properly reflect the impact of \pileup{} on the process of interest, and they can be more energetic than the process of interest, nevertheless only the process of interest is analysed in detail when using simulated samples.

% End of text imported from the .//sections/Datasets.tex input file

%

%
\section{Object reconstruction}
\label{sec:ObjReco}

% The next lines are included from the .//sections/Objects.tex input file
Reconstructing jets from \pileup collisions relies heavily on the capabilities of charged particle (track) reconstruction and particle flow objects (PFOs). First, good track reconstruction performance is instrumental in reconstructing different \pileup collisions as distinct clusters of tracks (\pvs) in a given \rbc. Furthermore, electron and muon tracks need to be associated with a given primary vertex with a very high accuracy in order to correctly identify and remove the \tpv{} and thus obtain an unbiased \pileup dataset. Finally, these tracking capabilities are critical for the assignment and reconstruction of jets from different \pvs.

This section describes the reconstruction of standard detector signatures in ATLAS, namely tracks, particle showers in the calorimeters, and primary vertices. These are used as building blocks in the reconstruction of physics objects such as electrons, photons, muons and jets from all \pvs to obtain a global interpretation of a given \rbc.

The reconstructed dataset contains overlapping signatures from both the triggering event and \pileup collisions. The final analysis selections demonstrated to produce a trigger-unbiased dataset are further described in \Section{sec:EvSel}.

\subsection{Standard object reconstruction}

Tracks are reconstructed in the ID by fitting a trajectory model to tracker hits~\cite{IDTR-2021-01}. Tracks with $p_\mathrm{T}^\mathrm{track} > 500\,\MeV$ are used as inputs for an iterative \pv reconstruction procedure~\cite{PERF-2015-01}. The reconstructed \pvs are sorted by decreasing $\sum \left( p_\mathrm{T}^\mathrm{track} \right)^2$. Vertices that are considered in this analysis are required to have at least two associated tracks.

Particle showers are reconstructed as clusters of energy deposits in the calorimeters~\cite{PERF-2014-07}. These \topos are formed by a topological clustering algorithm that identifies seed cells which have an energy at least four times the noise threshold. The algorithm iteratively expands around the seed cell by including any additional cells with an energy at least twice the noise threshold. As a final stage, any remaining neighbouring cells are added.

PFOs are obtained by combining tracks and \topos. The algorithm attempts to match tracks to calorimeter clusters. The energy expected to be deposited by the charged particle is calculated and used to judge whether the \topo is compatible with the charged particle hypothesis, or whether additional neutral energy is present. The expected energy is then subtracted on a cell-by-cell basis from the matching \topos, and any remaining energy indicative of neutral particles is left to form additional \topos. The tracks are classified as charged PFOs, while the surviving \topos are referred to as neutral PFOs. Charged and neutral PFOs are used as inputs to jet reconstruction in the particle flow algorithm, described in detail in Ref.~\cite{PERF-2015-09}.

Electron candidates are obtained by combining the information from the ID and the electromagnetic calorimeter~\cite{EGAM-2018-01}. Electrons are required to satisfy $\pt>7\,\GeV$ and $ |\eta| < 2.5$. Two types of electron candidates are considered. \texttt{Loose} electrons are required to satisfy the medium likelihood-based identification working point~\cite{EGAM-2021-01} and are used for trigger matching and \tpv identification. The inputs to the likelihood include measurements from the tracking system, the calorimeter system, and quantities that combine both tracking and calorimeter information. \texttt{Tight} electrons, on the other hand, are required to pass the tight likelihood-based identification working point and a tight isolation requirement. \texttt{Tight} electrons are used for an overlap removal procedure with jets.

Photons are reconstructed from clusters in the electromagnetic calorimeter~\cite{EGAM-2018-01}. Photons are required to have $\pt>7\,\GeV$ and $ |\eta| < 2.37$, and are not considered if they fall within the transition region between the barrel and endcap calorimeters
($1.37 < |\eta| < 1.52$). They are required to satisfy the tight identification criteria for the lateral and longitudinal shower shape and the tight isolation working point~\cite{EGAM-2021-01}. In general, photons cannot be easily associated with a specific \pv to the required level of precision. Photons are only used in the procedure to remove overlaps with jets.

Muon candidates are reconstructed by combining the hits from the ID and the muon spectrometer~\cite{MUON-2018-03}. Muons are considered if they are successfully reconstructed via combined, stand-alone, or segment-tagged algorithms, satisfy minimal quality criteria and have $\pt>7\,\GeV$ and $ |\eta| < 2.5$. Similar to electrons, two kinds of muon candidates are defined, differing in their isolation requirements. \texttt{Loose} muons, which are used for TPV identification, are required to pass a track-based isolation requirement $p_\mathrm{T}^\mathrm{varcone30}/p_\mathrm{T} < 0.07$. This requirement calculates the ratio of the total transverse momentum of surrounding tracks within a cone of $\Delta R=0.3$ (excluding the muon's own momentum) and helps suppress muons with a second high-momentum track nearby. This criterion was found to help reduce differences in \tpv identification between electron and muon triggers, likely due to reducing an inherent difference between such triggers in regards to heavy-flavour hadron decays: muon decays are more likely to pass a muon trigger, while electron decays are unlikely to pass an electron trigger due to the large surrounding hadronic component. \texttt{Tight} muons, which are only used for overlap removal with jets, are required to satisfy a tight particle-flow-based isolation requirement.

The \texttt{loose} electron and muon \pt thresholds are set significantly below the single-lepton trigger threshold (26 \GeV) to maximize the rate of \tpv{} identification and removal described in Section~\ref{sec:EvSel}. The \texttt{loose} offline lepton collections are required to be matched to the corresponding object at the HLT via a $\Delta R < 0.1$ requirement.

\subsection{\Pileup jet reconstruction}

To reconstruct jets from \pileup collisions, the standard ATLAS jet reconstruction algorithm is extended to all \pvs{}. This procedure is described in detail in Ref.~\cite{JETM-2023-08} and is summarized below. The \antikt algorithm~\cite{Cacciari:2008gp} with a radius parameter of \(R=0.4\) is used to reconstruct jets using PFOs as inputs using the \textsc{FastJet}~\cite{Fastjet} software package. The algorithm is run once for each reconstructed \pv{}. Only charged PFOs within $|(z_0 - z_\mathrm{PV})\sin\theta| < 1\,\mathrm{mm}$ are used for each vertex, where $z_0$ is the longitudinal position of the track at the point of closest approach in the transverse plane and $z_\mathrm{PV}$ is the primary vertex position along the beam axis. The association is exclusive, meaning that if there are multiple vertices satisfying this criterion, the charged PFO is only considered for the vertex that minimizes the longitudinal separation. The same set of neutral PFOs, origin-corrected to point to the current primary vertex, is used inclusively across all vertex iterations, as neutral PFOs carry no particular vertex interpretation. Since the jet clustering is run independently for each vertex, several jets are typically reconstructed for a given region of the calorimeter, corresponding to the same neutral PFOs adjusted corresponding to different vertex interpretations.

Jets from \pileup collisions are calibrated following the standard ATLAS procedure. This consists of several MC-based corrections that correct the average jet response to the particle-level scale, followed by \insitu calibrations that account for differences between data and simulated events~\cite{JETM-2018-05}. Wherever applicable, the calibration is modified to take into account the current vertex under consideration (which may not be the vertex with the highest $\sum(p_\mathrm{T}^\mathrm{track})^2$ as in the standard ATLAS jet reconstruction), but is otherwise identical. In Ref.~\cite{JETM-2023-08} the calibration was found to produce compatible jet energy scale responses between different PVs in simulated events.

Due to its combinatorial nature, the modified jet reconstruction procedure produces a large number of redundant jets. The jet-vertex tagger (JVT)~\cite{PERF-2014-03} is used to select jets that are consistent with their vertex reconstruction hypothesis: charged tracks pointing at the jet must primarily originate from the vertex in question, and the ratio of the jet's charged activity to its total energy must roughly correspond to hadronic showering expectations. This significantly reduces the number of jets that do not have significant charged activity associated with the current vertex of interest.

\Pileup jets that satisfy $\pt>17\,\GeV$, $|\eta| < 2.0$ and the JVT requirement are retained for analysis.
Overlaps between jets and other physics candidate objects are considered.
Any jet found within $\Delta R < 0.2$ of a \texttt{tight} electron is removed.
Any jet with less than three tracks associated to it found within $\Delta R < 0.2$ of a \texttt{tight} muon, or if a muon inner-detector track would be clustered into the jet, is removed. Finally, any jet found within $\Delta R< 0.4$ of a photon is removed.

These requirements impose a stringent vertex interpretation for most \pileup jets, but uncommon cases of overlapping showers from differing \pvs{} remain. The procedure for addressing these jet--jet overlaps is described in \Section{sec:EvSel}.

% End of text imported from the .//sections/Objects.tex input file

%

%
\section{Analysis selections}
\label{sec:EvSel}

% The next lines are included from the .//sections/EvSel.tex input file
The final dataset used in this search is obtained after a series of sequential selections applied on the reconstructed \pileup data, first at the object level as discussed in \Section{sec:ObjReco}, then at the \rbc level and finally at the \pv level, closely following the selections used in Ref.~\cite{JETM-2023-08}.

\rbcs are required to pass either the single-electron or single-muon trigger requirement, and a veto is applied for \rbcs with errors identified in the Tile, LAr or SCT detectors~\cite{DAPR-2018-01,LARG-2013-01}.
In addition, \rbcs with jets arising from sources other than \ppp{} collisions are removed via Loose jet cleaning criteria~\cite{ATLAS-CONF-2015-029}. %
It is required that a \tpv can be found, as described in detail in Ref.~\cite{JETM-2023-08}, by matching an offline loose lepton to its triggering counterpart and requiring that the offline lepton can be uniquely associated to a \pv.
If either of these conditions fails, and the collision of origin for the triggering lepton cannot be found, the entire \rbc is vetoed.

The \pvs in the surviving \rbcs are examined next.
First, in order to ensure that the analysed \ppvs in the \rbc remain unbiased by the trigger, the TPV is vetoed together with any \pvs within $\left|\Delta z(\mathrm{TPV,PV})\right|<2\mathrm{\ mm}$.
The remaining \ppvs are required to have at least two jets with $\pt > 17\,\GeV$ and $|\eta|<2.0$ matched to them; these two leading jets form the dijet system used in the analysis.
The robustness of the \tpv{} subtraction procedure is checked by comparing the dijet invariant mass spectrum between \rbcs selected by electron and muon triggers; the two spectra were found to be compatible within the statistical precision of the dataset.

In order to reduce contributions from overlapping high-momentum signals originating in different \ppp{} collisions, the two leading jets for each \ppv are required not to overlap with other jets from a different \ppv within the \rbc.
This requirement is achieved by means of a geometrical isolation criterion: jets from different \ppvs are considered to overlap if they are within $\Delta R=0.4$ of each other.
The isolation requirement exceptionally lowers the jet \pT{} threshold from 17~\GeV{} to 15~\GeV{} and increases the detector acceptance from $\left|\eta\right|<2.0$ to $\left|\eta\right|<2.4$.
This conservative choice is made to ensure that jets are not mistakenly qualified as isolated because the jet they overlap with happens to be just past the kinematic or geometric threshold used, thereby better capturing stochastic overlaps near analysis thresholds.
If such an overlap is identified, only the affected \ppv is removed from the analysis, while other \ppvs in the same \rbc\ that satisfy the isolation requirement are retained.
More details on this isolation criterion can be found in Ref.~\cite{JETM-2023-08}. An additional jet cleaning selection is imposed to reject \ppvs in which either of the two leading jets deposits more than half of its energy in the Tile Calorimeter gap and cryostat scintillators, located in a narrow Tile gap region from $1.0 < |\eta| < 1.6$. Such jets are susceptible to large calibration corrections and noise fluctuations due to the limited active material in this region, and were observed to degrade the background modelling.
Additionally, a requirement that the timing\footnote{The jet timing is calculated as the energy-squared-weighted sum of the timing of the constituent clusters, which in turn is the energy-squared-weighted sum of the timing of the cells within each cluster.} of the two leading jets be within 10~ns of the nominal bunch-crossing time is applied in order to reduce contributions from out-of-time \pileup.
Finally, the two leading jets in each \ppv are required to satisfy the condition $\left|y^*\right| =\frac{\left|y_1-y_2\right|}{2} < 0.8$, where $y_1$ ($y_2$) is the rapidity of the leading (subleading) jet. This selection enhances the presence of resonant signal-like topologies while suppressing contributions from the main background, SM multijet processes, which increases with $\left|y^*\right|$ due to $t$-channel poles at leading order. The threshold was found to be the optimal selection for maximal Poisson significance ($S/\sqrt{B}$) for the models targeted in this search, estimated with $Z'$ simulated data samples for resonance masses between 80 and 250~GeV.

A unique aspect of this analysis is that it is possible for a single \rbc{} to have multiple \ppvs{} passing all analysis selections, although it is rare in practice~\cite{JETM-2023-08}.
In such cases, each \ppv{} is treated as a separate event, as the by-product of each \ppp{} collision is an independent physical process.
Concretely, this means that a single \rbc{} can contribute multiple entries to any distribution, including the dijet invariant mass.

\begin{figure}[h]
\centering
\includegraphics[width=0.7\textwidth]{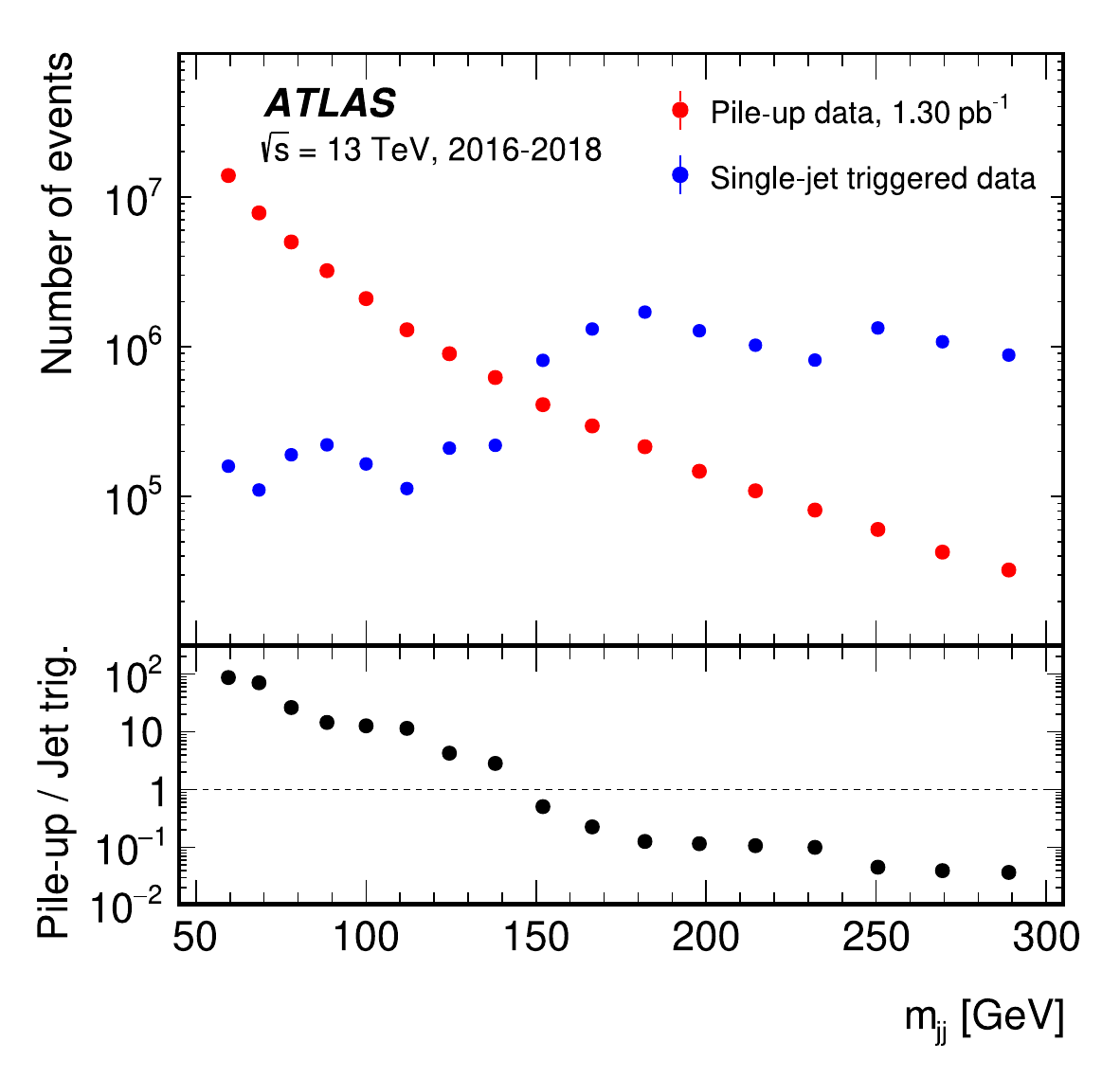}
\caption{Dijet \mjj distribution in the pile-up dataset compared to the single-jet triggered dataset. In both cases the full analysis selections are applied, except for the requirements related to the removal of the triggering \ppp{} collision in the case of the single-jet triggered data.
The single-jet triggered spectrum is made up of several  prescaled triggers, with increasing jet kinematic thresholds, which are stitched together.
To emulate what a real analysis would do, each bin corresponds to a single trigger, where the trigger with the most events that is also fully efficient (post-turn-on regime) is used.
The bottom panel shows the ratio of the two distributions.
Statistical uncertainties are included, and are negligible}
\label{fig:StatsComparison}
\end{figure}

The acceptance times efficiency of each of these selections is summarised in Table~\ref{tab:Efficiencies}.
The efficiencies are split between data-based effects, which relate to the \pileup{} dataset reconstruction procedure independent of a specific process of interest, and simulation-based effects, which are evaluated for three mass hypotheses for the $Z'$ model.
The data-based efficiencies are derived following the procedure described in Ref~\cite{JETM-2023-08}.
A comparison of the statistical power of the pile-up reconstruction dataset to the standard single-jet triggers used during ATLAS data taking~\cite{TRIG-2016-01} is shown in Figure~\ref{fig:StatsComparison}, which includes the full analysis selections for both cases. The standard jet triggers are prescaled, meaning that events that would pass the trigger are only recorded a fraction of the time; this is necessary to keep the trigger rate under control and within the trigger system capabilities, as otherwise single-jet triggers would record essentially every \ppp{} bunch crossing.
The pile-up dataset has a higher statistical power compared with the standard prescaled single-jet triggers up to 140~GeV in the invariant mass of the dijet system, with a factor of 100 larger statistics at the beginning of the explored spectrum.\footnote{For this comparison only the size of the collected dataset is taken into account. A dijet search using prescaled single-jet-triggered data would not be able to fully leverage the shown statistical power, as the different triggers do not represent a single dataset. A real dijet search would also need to anchor the background fit at the start of the spectrum, thus the search phase would in reality start later, and the pile-up dataset would have leading statistical power up to higher masses than shown.}

\begin{table}
\centering
\caption{Acceptance times efficiency for the sequential selections applied in the analysis for three benchmark $Z'$ mass hypotheses with $g_q=0.6$. The efficiency for the TPV identification and removal of vertices that are close by in $\Delta z$, on the first two rows, are computed from data, since these do not depend on the dijet topology targeted. The TPV identification efficiency includes losses from the triggering lepton reconstruction step and matching to a \pv. The efficiency for the removal of the TPV and nearby vertices only depends on the geometry of the bunch crossing, and is therefore estimated as the average efficiency of vertices surviving this step for the entire dataset. The acceptance times efficiency for the rest of the selections is estimated from MC simulated event samples for each of the signal hypotheses. The final row shows the total efficiency times acceptance entering the analysis for each signal hypothesis, which is obtained for each column as the product of all rows together.}
\begin{tabular}{c|c|c|c}
Selection & $m(Z') = 100\,\gev$ & $m(Z') = 150\,\gev$ & $m(Z') = 250\,\gev$\\
\hline
\multicolumn{1}{l|}{\textbf{Data-based}} & \multicolumn{3}{c}{ }\\
TPV identification & \multicolumn{3}{c}{87.5\%}\\
$\left|\Delta z(\mathrm{TPV,PV})\right|>2\mathrm{\ mm}$ & \multicolumn{3}{c}{92.2\%}\\
\hline
\multicolumn{1}{l|}{\textbf{MC-based}} & &&\\
$\geq$2 jets ($\pt > 17\,\gev$, $|\eta|<2$) & 39.7\% & 48.3\% & 58.3\%\\
Overlapping jet, timing, and Tile gap criteria & 78.3\% & 85.5\% & 91.7\%\\
$|y^*|<0.8$ & 82.2\% & 81.0\% & 80.1\%\\
\hline
\textbf{Total efficiency times acceptance} & 20.6\% & 27.0\% & 34.5\%
\end{tabular}
\label{tab:Efficiencies}
\end{table}

% End of text imported from the .//sections/EvSel.tex input file

%

%
\section{Methodology}
\label{sec:Methodology}

% The next lines are included from the .//sections/BkgEstimate.tex input file
\subsection{Resonance search} \label{sec:resoSearch}
The invariant mass range explored in this analysis is dominated by QCD multijet background, which is modelled using a smoothly-falling function, with small additional expected resonant contributions from hadronic decays of SM $W$ and $Z$ bosons. Inclusive observations of the hadronic $W/Z$ dijet signal at a hadron collider were previously reported by the UA2 Collaboration~\cite{Ansari:1987452,Alitti:210698}. Producing simulated event samples for inclusive $W$ and $Z$ bosons with hadronic decays, possibly in association with other jets, and at NLO accuracy with state-of-the-art MC generators, was found to be challenging due to divergences in collinear and infrared regimes. Instead, these resonances are modelled with Gaussian templates, justified since the detector resolution of the $W$ and $Z$ resonances exceeds their intrinsic resonance widths.
The final template used, $f_{W/Z}$, is the sum of two Gaussians, one centred at $m_W=80$~GeV and the second one at $m_Z=91$~GeV.
A value of $\sigma = 15\mathrm{\ GeV}$ is considered for both resonances, estimated from the simulated MC event sample for a $Z'$ boson with mass of 80~GeV, as the experimental resolution does not change significantly over that mass range.

An overall normalisation factor is included for both Gaussians, with a relative normalisation factor applied to the template for the $Z$ boson to account for the difference in cross section and branching ratio to quarks between both resonances~\cite{STDM-2015-03}. The lack of a precise estimate limits the ability to disentangle potential BSM signals near the $W/Z$ resonance, and therefore the BSM interpretations of the search are only performed for masses above 100~GeV.

The first step in the analysis is to study the compatibility of the data with the background-only hypothesis, arising from SM multijet processes (Section~\ref{sec:BkgEstimate}), where the procedure for quantifying the degree of compatibility must be defined in advance of looking at the data. If the $\chi^2$ $p$-value of the background-only fit is smaller than 0.05, the $W/Z$ template is added to the background model to account for the contribution of hadronic SM $W/Z$ processes. Due to being near the start of the considered \mjj spectrum, any SM $W/Z$ contribution may either show up as a local excess at $m_{W/Z}$ or may bias the start of the fit, thereby creating tension later in the \mjj spectrum. The spectrum is then refit using this new background model.

If the $\chi^2$ $p$-value is still smaller than 0.05, the \texttt{BumpHunter} algorithm~\cite{CDF:2008voc,Choudalakis:2011qn} is subsequently used to identify local excesses as a proxy for resonance-like behaviour present in the data. It performs a scan of all possible adjacent windows in the fit mass range, and evaluates the consistency of the observed yields with Poisson fluctuations over the expected number of events.
The background-only fit is repeated after removing the highest local excess identified by \texttt{BumpHunter}, where a new $\chi^2$ $p$-value of 0.05 or higher indicates compatibility with the hypothesis of a single resonance on top of the smoothly falling background.
In this case, there remains an ambiguity: the original fit tension may have been related to the SM $W/Z$ contribution being large enough to affect the background model description, the presence of a local excess, or a combination of both effects.
The relevance of the SM $W/Z$ contribution is therefore tested by removing the $W/Z$ template from the background model and retaining the same \texttt{BumpHunter} window exclusion; in case the $\chi^2$ $p$-value remains above 0.05, the $W/Z$ template is dropped from the background model to simplify the background description, otherwise it is retained.
After the inclusion or exclusion of the $W/Z$ template in the background model has been fixed, the \texttt{BumpHunter} excess is studied in the context of the $Z'$ scenario discussed in Section~\ref{sec:DataMC}.

If the fit of the background function to the data yields a $\chi^2$ $p$-value of 0.05 or higher, the background-only hypothesis is not rejected. In this case, exclusion upper limits on the cross section ($\sigma$) times acceptance times efficiency ($A$) times branching ratio to jets ($\mathrm{B}$), $\sigma\times A\times \mathrm{B}$, are set for the SM and BSM resonance interpretations. The upper limits are set for the $W/Z$ resonance, described as a double-Gaussian template, as well as for two signal scenarios: the $Z'$ models discussed in Section~\ref{sec:DataMC} and Gaussian-like dijet resonances for masses above 100 GeV. Only signals with resonance masses between 100 and 250~GeV are considered in the upper limit computation, in order to ensure that the majority of the resonance shape is contained within the fit range used for the background estimate, introduced in detail in Section~\ref{sec:BkgEstimate}.

Templates for benchmark points in the $Z'$ scenario are generated with MC simulation, as discussed in Section~\ref{sec:DataMC}. The signal templates are fitted with a double-sided Crystal Ball (DSCB) function~\cite{HIGG-2014-04}, and the resulting parameterization is subsequently employed in the signal-plus-background fits.
The mean of the Gaussian bulk was found to be shifted downwards by around 6--7\% compared to the generated $Z'$ mass.
The granularity of the mass grid for a given coupling $g_q$ is increased with additional templates created by performing an interpolation between the available event samples, following a procedure similar to the one discussed in Ref.~\cite{Baak:2014fta}. For the Gaussian signal scenarios, templates built with a Gaussian distribution with mean equal to the resonance mass are used. In this case, the relative width of the distribution, representing the total width including both the intrinsic resonance width and the detector resolution, is considered at values of 10\%, 15\%, 20\%, and 25\% of the resonance mass. As the relative detector mass resolution improves with increasing mass, the lower bound of 10\% is chosen to correspond to the resolution at the highest mass point considered. The upper bound of $25\%$ is fixed by fit considerations, as above this width fitting the signal component becomes increasingly more difficult and thus the signal sensitivity starts to drop rapidly.

\subsection{Background estimate}\label{sec:BkgEstimate}

The non-resonant background in this search corresponds to SM multijet processes.
Due to the large statistics in the \pileup dataset in the targeted mass range, producing simulated samples for multijet processes with enough statistical power is computationally unfeasible.
Instead, the invariant mass of the two leading jets of the selected \ppvs, \mjj, is fit with a family of functions that preserve the smoothly falling behaviour expected from SM multijet processes:
\begin{equation}
f_N(x) = p_1(1-x)^{p_2}x^{\sum_{i=3}^Np_i\log^{(i-3)}(x)},
\label{eq:dijet}
\end{equation}
where $x\equiv\mjj/\sqrt{s}$ and $N$ is a fixed number of parameters.
Such functions were found to provide appropriate background estimates in a number of dijet and multijet resonance searches by the ATLAS and CMS collaborations~\cite{EXOT-2010-01,EXOT-2010-07,EXOT-2013-11,CMS-EXO-10-010,CMS-EXO-11-015,CMS-EXO-19-012,EXOT-2018-05,EXOT-2018-39}.
In this analysis, versions of the functional form with three to nine free parameters were tested.

An additional turn-on term is included in the description to account for deviations from the behaviour described by Equation~\ref{eq:dijet} at the beginning of the \mjj spectrum due to jet \pt threshold effects and reconstruction inefficiencies.
The chosen functional form to fit the data is thus
\begin{equation} \label{eq:turnOn}
f_\textrm{fit}(x, N) = \left(\frac{1}{1+e^\frac{-(x-x_0)}{k}}\right)f_N(x),
\end{equation}
where $x_0$ is the point where the turn-on function reaches 50\% efficiency, and the parameter $k$ controls the steepness of the turn-on curve. Both the $x_0$ and $k$ parameters are allowed to float freely in the fit.
The data are divided into narrow 0.5~GeV equidistant bins as an approximation to an unbinned fitting procedure. This bin width is much smaller than the detector mass resolution, which ranges from approximately 12 to 18~GeV over the mass range relevant to the analysis.
The fit is then performed using a framework based on the \texttt{RooFit}~\cite{RooFit} and \texttt{RooStats}~\cite{RooStats} codes.
The \mjj range used in the fit spans from 55 to 300~GeV.
After the fit, the data are rebinned into mass-dependent bin sizes that reflect the detector resolution, as estimated with dijet MC simulated event samples.
The rebinned data are also used as input in deriving the fit $p$-values in the \texttt{BumpHunter} algorithm.

A number of cross-checks are performed on the data, including spurious signal and signal injection tests, to ensure the robustness of the fitting procedure for fixed $N$ in describing the background and its ability to mask potential signals present in the data following the recommendations given in Ref.~\cite{ATL-PHYS-PUB-2020-028}.
First, pseudodata are created by fitting the data with a $f_\mathrm{fit}(x,N+1)$ function and sampling from it the \mjj distribution multiple times, applying Poisson fluctuations that match the statistical uncertainties observed in data.
The fit with $f_\mathrm{fit}(x,N+1)$ is required to have a $\chi^2$ $p$-value higher than 0.05 to ensure the compatibility of the data with the background-only hypothesis.
This ensures the pseudodata have realistic statistical power while containing no resonant structure.

In the spurious signal tests, signal-plus-background fits are performed to the pseudodata distributions, where the background component is given by $f_\textrm{fit}(x, N)$.
Two options for the signal hypothesis are used:
Gaussian-like resonance templates under different mass hypotheses within the fit range and widths between 10 and 25\% of the mass  provide model-independent evaluation of bias in the signal extraction, and simulated $Z'$ signal templates under the different mass and coupling $g_q$ hypotheses discussed in Section~\ref{sec:DataMC}.
The spurious signal test checks whether the extracted amount of signal is compatible with the background-only distribution used to generate the pseudodata, revealing potential biases in the fitting procedure.
For a given signal hypothesis, a signal yield is extracted from each pseudodata distribution.
The mean $S_\text{spur}$ and standard deviation $\sigma_\text{spur}$ of the distribution of signal yields are required to fulfil the condition $S_\text{spur} / \sigma_\text{spur} < 0.5$ for all signal hypotheses.

In the signal injection test, signal templates are added to the pseudodata distributions to evaluate whether the amount of signal present in the data can be extracted without bias.
Both Gaussian shapes and $Z'$ signal templates are used, as well as the $W/Z$ double Gaussian template, injected with a given cross-section such that the significance $S/\sqrt{B}$ is in the range 0-5.
The number of signal $S$ and background $B$ events are estimated in each test as the contribution of each process to a full-width at half-maximum (FWHM) window around the peak of the injected signal.
The mean $S_\mathrm{inj}$ of the extracted signal yield distributions is required to be within 0.5$\sigma_\mathrm{inj}$ of the injected signal yield for all signal hypotheses, with $\sigma_\mathrm{inj}$ being the standard deviation of the extracted signal yield distribution.

The final fitting procedure applied to the full dataset uses a function of the form $f_\textrm{fit}(x, N=5)$ to describe the observed \mjj spectrum in the \pileup dataset.
This number was chosen by performing an $F$-test~\cite{ATL-PHYS-PUB-2020-028} to the functions that pass all spurious signal and signal injection tests in a statistically reduced dataset of about one fifth of the total luminosity of the \pileup data, constructed by randomly selecting one in every five \rbcs from the dataset with the requirements discussed in Section~\ref{sec:EvSel}.
This was done to ensure the chosen function would be optimal when describing a dataset with negligible signal contribution, containing mostly background events.
The validity of this choice to use a sub-dataset to characterise the fit quality was studied by repeating the same set of tests using a roughly $1/35^\mathrm{th}$ dataset, where the results were cross-checked on the one-fifth dataset.

% End of text imported from the .//sections/BkgEstimate.tex input file

%

%
\section{Systematic uncertainties}
\label{sec:Uncertainties}

% The next lines are included from the .//sections/Systematics.tex input file
A number of experimental sources of uncertainty are considered in the interpretation of the analysis under different signal hypotheses.
Systematic uncertainties in the jet energy scale and resolution are evaluated on a jet-by-jet basis on the simulated $Z'$ samples discussed in \Section{sec:DataMC}.
An ensemble of 43 nuisance parameters are considered, following the description in Ref.~\cite{JETM-2018-05}.
The impact on the signal templates is estimated by varying each source of uncertainty up and down by one standard deviation.
The nominal and varied \mjj distributions are fitted with a DSCB function, and the impact of each systematic variation on the mean and width of the distribution is applied as a systematic uncertainty.
The largest jet-energy-related sources of uncertainty on the mean of the signal are the jet flavour composition and response, with differences up to 2~GeV.
The impact of the width of the signal templates is negligible, with the largest contributions equally coming from the jet flavour composition and response, which remain at the sub-GeV level.

An uncertainty on the procedure used to adjust the simulated \pileup{} conditions to match those in recorded data, as applied to the simulated $Z'$ samples, is derived by varying the total inelastic cross-section used in the determination of the distribution of the number of expected collisions in simulation. The resulting alternative weights are applied to simulated events and the effect on the final observables is taken as a systematic uncertainty.

Uncertainties in the signal yield extraction due to the background modelling are accounted for by including the amount of spurious signal as part of the systematic uncertainty prescription.
For each signal hypothesis, a separate signal component with equal shape is implemented, with maximum amplitude $0.5\sigma_\mathrm{spur}$ and constrained by Gaussian terms in the likelihood. %

As discussed in Ref.~\cite{JETM-2023-08}, the uncertainty on
the luminosity of the collected \pileup dataset is 0.83\%.
It is applied as an uncertainty on the normalisation of the extracted signals from the fit, and is found to be negligible with respect to the other sources of experimental uncertainty.

Electrons and muons are important to the reconstruction of the \pileup{} dataset: they are the means by which TPV(s) are identified and thus how trigger bias is removed.
This is of relevance solely to data, where the impact of using a triggered dataset must be mitigated; in simulation, there is no such concern.
Typical lepton uncertainties related to differences in their reconstruction and identification between data and simulation are thus not applied, as leptons are never used in simulation.
The efficiencies related to the use of leptons in the TPV removal process, namely lepton reconstruction, offline-online matching, and vertex identification, are estimated directly in data using the full analysis dataset and the uncertainties on those efficiencies are found to be negligible.
This factorisation assumes that vertices are sufficiently uncorrelated so that the jets used in the analysis are not impacted by the presence (or absence) of a TPV, which was demonstrated to be a valid assumption in Ref.~\cite{JETM-2023-08}.

%
%
%
%

% End of text imported from the .//sections/Systematics.tex input file

%

%
\section{Results and exclusion limits}
\label{sec:result}

\begin{figure}[h!]
\centering
\includegraphics[width=0.6\textwidth]{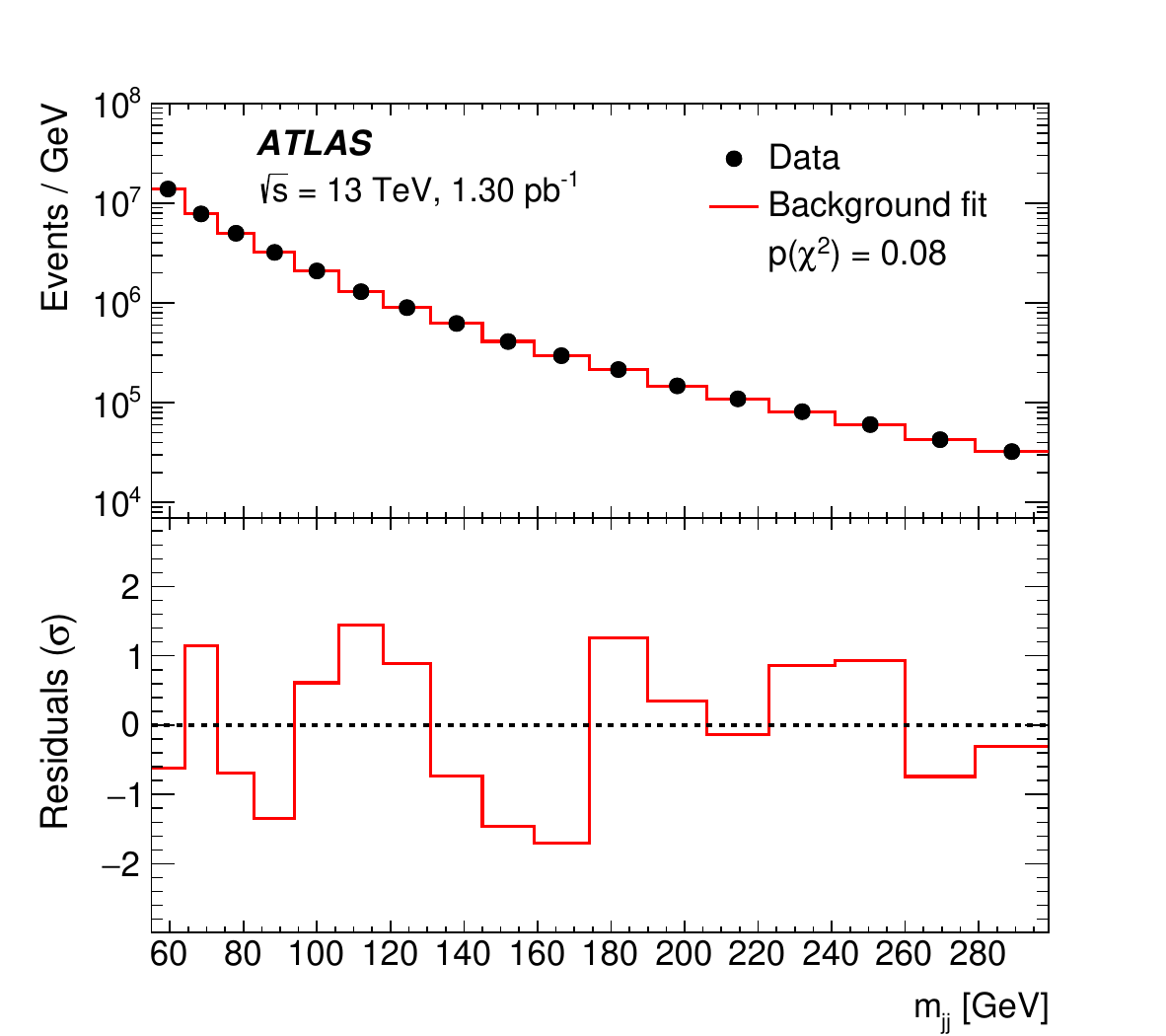}
\caption{
Dijet \mjj distribution in data (black markers) shown in variable-width bins corresponding to the experimental resolution after all analysis selections discussed in the text. The histogram in the upper panel shows the background estimate after the background-only fit. The bottom panel shows the residuals between data and the fitted background estimate in terms of standard deviations ($\sigma$), where $\sigma$ corresponds to the data statistical uncertainty.
}
\label{fig:mjj}
\end{figure}

% The next lines are included from the .//sections/Results.tex input file
The \mjj distribution for \ppvs in data that survive the selections discussed in Section~\ref{sec:EvSel} is shown in Figure~\ref{fig:mjj}, in variable-width bins corresponding to the \mjj resolution obtained from simulated dijet event samples.
The resulting background estimate function $f_\mathrm{fit}(x, N=5)$ after the fit is shown together with the data, as are the corresponding residuals of the fit, which are compared to the data statistical uncertainty.

The fit function was chosen based on studies using a reduced dataset, as described in Section~\ref{sec:BkgEstimate}.
The fitted template describes the data well, with a $\chi^2$ $p$-value of 0.08; consequently, the background-only hypothesis is not rejected. As explained in Section~\ref{sec:resoSearch}, $p$-values greater than 0.05 also indicate that the background modelling is sufficient without the inclusion of the $W/Z$ template, which is therefore not included in the estimate. The fitted turn-on parameters, as defined in Equation~\ref{eq:turnOn}, are found to be $x_0 = 43.7 \pm 0.2$~GeV and $k = 5.5 \pm 0.1$. Without the inclusion of the sigmoid turn-on curve, the $\chi^2$ $p$-value is 0, and the residuals reach up to 12 standard deviations; the sigmoid is therefore an important part of properly describing the background distribution.

When using the \texttt{BumpHunter} algorithm, the window width is allowed to span between one and six resolution bins. The largest excess is found between 106 and 131~GeV, with a local significance of 1.7$\sigma$. The corresponding global $p$-value of 0.67 indicates no statistically significant deviation from the background expectation.

Given no significant excess over the background prediction is found, the background model is accepted.
The analysis proceeds to search for two signal scenarios: $Z^\prime$ signals, as discussed in Section~\ref{sec:DataMC}, and Gaussian-like dijet resonances, both for masses above 100~GeV.
Upper limits on $\sigma\times A\times B$, the product of the cross section, acceptance times efficiency, and branching ratio to jets, are then set for both of these signal scenarios at 95\% CL.

Signal-plus-background fits are performed for the $W/Z$ as well as for each individual hypothesis for both signal scenarios.
For the $Z'$ and Gaussian signals, the systematic uncertainties discussed in Section~\ref{sec:Uncertainties} are introduced as nuisance parameters constrained with Gaussian terms in the likelihood. For the $W/Z$ interpretation, the corresponding spurious signal systematic uncertainties are included in the fit and propagated to the derived cross section limits.

Similarly to the procedure followed in Ref.~\cite{EXOT-2018-39}, the parameters related to the background estimate are left to float in the signal-plus-background fits to take into account the functional form flexibility and its interplay with the signal template considered.
A modified frequentist approach \cite{Cowan:2010js} is used to derive upper limits at 95\% CL with the $\mathrm{CL}_\mathrm{S}$~\cite{Read:2002hq} technique for each signal scenario.
The best fit for the $W/Z$ template normalisation is consistent with zero, thus the analysis does not have sensitivity to observe the $W/Z$ signal.
The expected (assuming no $W/Z$ signal) and observed 95\% confidence level upper limits on the yield of the $W$ and $Z$ boson template are:

\begin{equation}
N_{\text{exp}}^{95\%\,\text{CL}} = (7.1^{+2.8}_{-2.0})\times 10^{4}, \quad N_{\text{obs}}^{95\%\,\text{CL}} = 5.8 \times 10^{4}.
\end{equation}

These limits are compared to the expected contribution from SM $W$ and $Z$ bosons to the analysis after selections. The production cross sections for each process as measured by ATLAS are $\sigma(pp\rightarrow W)=190.1\mathrm{\ nb}$ and $\sigma(pp\rightarrow Z)=60.18\mathrm{\ nb}$~\cite{STDM-2018-17}, which for the luminosity of the pile-up dataset, and after considering the acceptance of the analysis selections in the relevant mass region, yields a total expected contribution of $3.9\times10^4$~events. While the SM $W$ and $Z$ bosons are not observed in this result, the expected exclusion limit of $1.8\times\sigma_\mathrm{SM}^{W/Z}$ is at a level that requires investigating these processes.

It is possible that hadronic $W$ and $Z$ boson decays are reconstructed at different masses than the values assumed in the $W/Z$ template. Both theoretical and experimental effects can shift the peak position: for example, interference between the QCD multijet background and $W/Z$ production process can lead to few-GeV effects~\cite{RANFT1979122,BAUR1989519,Alitti:210698,Ding:2019bkd}, and various jet-related reconstruction effects can also shift the mass peak.
Due to the very large width of the template, such shifts in the peak position are not expected to have a significant impact on the result.
In order to confirm this, a fit using an alternative $W/Z$ template obtained by shifting the mean of both Gaussians down by 7\% is performed to investigate the impact of a possible downward shift of the reconstructed peak positions; this is done following the results seen using $Z'$ simulated samples as mentioned in Section~\ref{sec:resoSearch}. The resulting limits for the shifted template are found to be compatible with the nominal results. This cross-check also covers the impact of interference effects between the QCD multi-jet background and $W/Z$ production, which are expected to result in a downward shift of a few GeV.

\begin{figure}[h]
\centering
\includegraphics[width=0.6\textwidth]{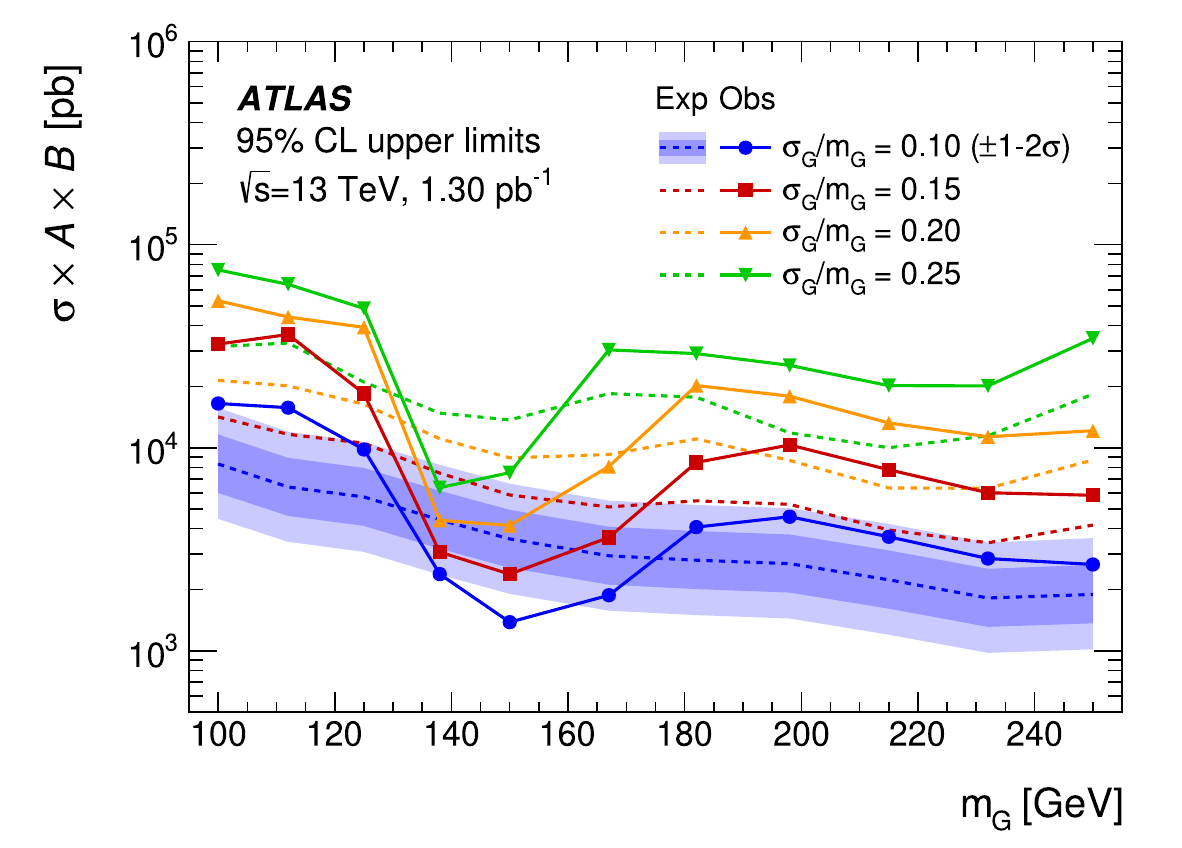}
\caption{Expected (dashed lines) and observed (solid lines) 95\% CL upper limits on $\sigma\times A\times B$ for Gaussian signal templates as a function of the resonance mass, $m_G$, for relative widths of $\sigma_\mathrm{G}/m_\mathrm{G}=10\%$, 15\%, 20\% and 25\%. For the Gaussian model with 10\% width, the shaded bands indicate the $1 \sigma$ and $2 \sigma$ uncertainty ranges around the expected limit.}
\label{fig:upperLimits_Gaussian}
\end{figure}

The lack of evidence for the presence of the $W/Z$ mass peak results in a poorly constrained $W/Z$ template, which when combined with a BSM signal shape leads to fit instability.
The $W/Z$ template is therefore not included in the background fit when conducting BSM signal interpretations.
The $\sigma\times A\times B$ limits on Gaussian resonance scenarios are shown in Figure~\ref{fig:upperLimits_Gaussian}.
The results are linearly interpolated among the tested benchmark mass hypotheses.
The observed limits are consistent with the expectation from the background-only case.
The fluctuations in the observed limits are strongly correlated with the residuals observed in Figure~\ref{fig:mjj}.

\begin{figure}[h]
\centering
\includegraphics[width=0.6\textwidth]{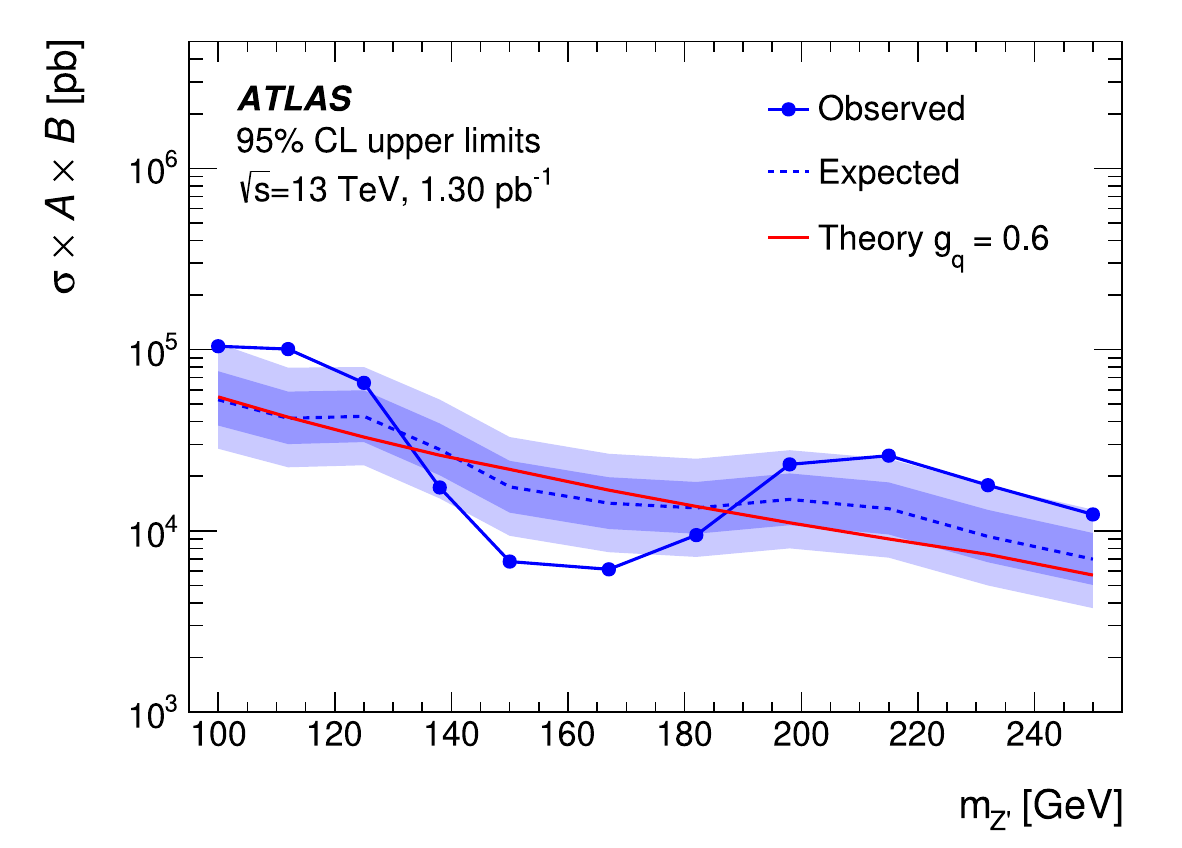}
\caption{Observed (solid lines) and expected (dashed lines) 95\% $\mathrm{CL}$ upper limits on $\sigma\times A\times B$ for $Z'$ signal templates with $g_q=0.6$ coupling. The shaded bands around the expected limit indicate the $1 \sigma$ and $2 \sigma$ uncertainty ranges. The theory line represents the expected $\sigma\times A\times B$ estimated from the generated signal samples.}
\label{fig:mjj-xsec-limits}
\end{figure}

Figure~\ref{fig:mjj-xsec-limits} shows the exclusion limits on $\sigma\times A\times B$ for the $Z'$ axial-vector mediator with $g_q = 0.6$ coupling.
The expected exclusion limits sit very close to the production cross-section line, and the observed limits exclude masses between 135 and 185 GeV for this coupling hypothesis.
The most significant localized excess is found for a $Z'$ signal with a mass of $m_{Z'}=112$~GeV for $g_q$ of either 0.6 or 0.8, corresponding to a local significance of 3.0$\sigma$.
Accounting for the look-elsewhere effect, this corresponds to a global significance of 2.3$\sigma$, evaluated using dedicated pseudodata samples derived from the final background-only estimate.

The global $\mathrm{CL}_\mathrm{S}$ $p$-value is distinct from the global \texttt{BumpHunter} $p$-value, as the former tests a specific signal-plus-background hypothesis against the background-only hypothesis at each mass point, while the latter scans for shape-independent localized excesses across all possible intervals by comparing the data with the bin-by-bin estimate from the background-only hypothesis.
The \texttt{BumpHunter} test therefore indicates that the distribution of observed events is compatible with a background fluctuation with the obtained $p$-value of 0.67,
while the $\mathrm{CL}_\mathrm{S}$ result indicates that the observed data at $m_{Z'} = 112$~GeV is more consistent with the signal-plus-background hypothesis at that mass than with the background-only hypothesis, but there is insufficient evidence to reject the background-only hypothesis.

\begin{figure}[h]
\centering
\includegraphics[width=0.7\textwidth]{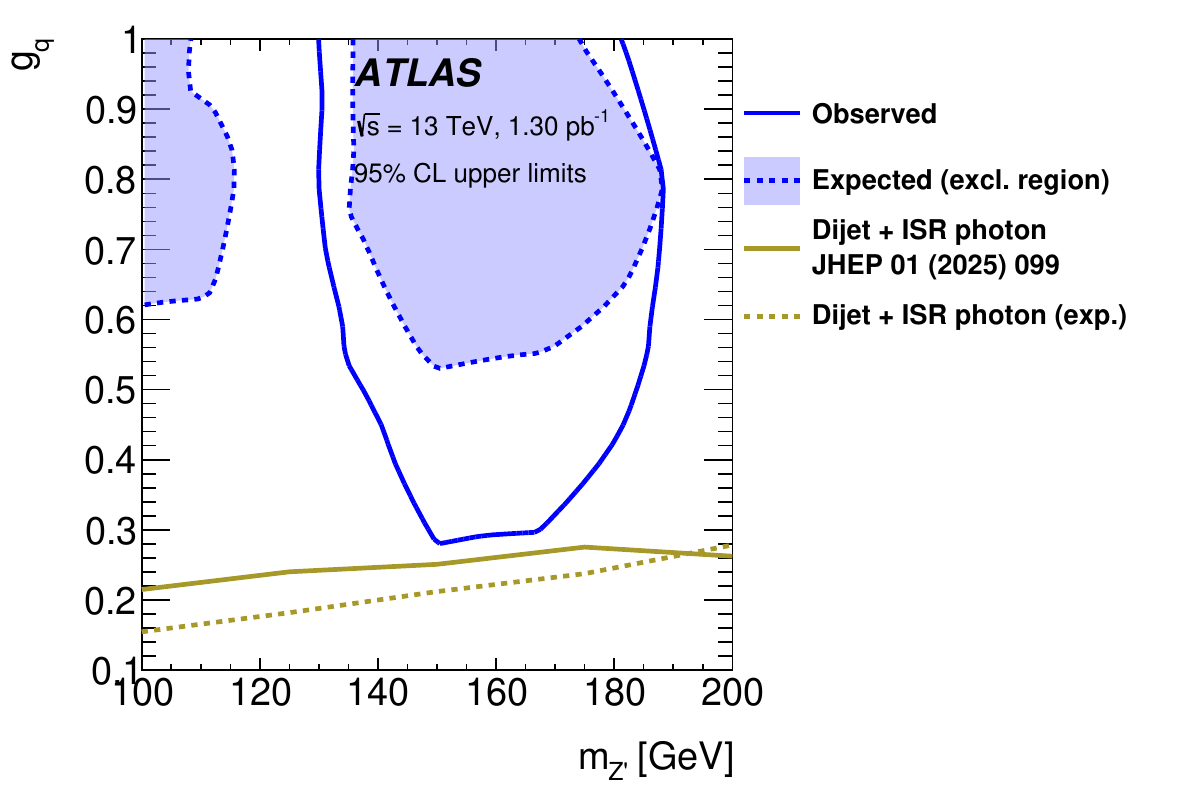}
\caption{Observed and expected 95\% $\mathrm{CL}$ upper limits on the $g_q$ coupling as a function of the resonance mass.}
\label{fig:mjj-gq-limits}
\end{figure}

Using all of the available $Z'$ samples, cross-section limits are translated into exclusion limits on the $g_q$ coupling, which are shown in Figure~\ref{fig:mjj-gq-limits}.
Mass hypotheses between 135 and 185~GeV are excluded for couplings $g_q=0.6$, with the lowest value of $g_q$ excluded being 0.28 for $m_{Z'}=150$~GeV.
The shape of the exclusion regions on this plot is explained by two main factors.
First, as seen in Figure~\ref{fig:mjj-xsec-limits}, both the expected and observed limits oscillate, which is a common residual effect seen in dijet resonance searches using functional-form background estimates; in this case it happens that the oscillation crosses the line defining the theoretical expectation.
Second, as $g_q$ increases, the $Z^\prime$ width increases, thereby spreading the signal out over more bins; at some point the width gets so large that the sensitivity degrades faster than the production cross-section increases.
The first effect explains the structure of the limits as a function of the resonance mass, while the second explains why the excluded regions do not significantly grow as $g_q$ is increased.

Figure~\ref{fig:mjj-gq-limits} also shows the corresponding $g_q$ exclusion limits from the boosted dijet+ISR photon~\cite{EXOT-2022-33} search. While more stringent limits are obtained for the $Z'$ model using boosted techniques, the pile-up-based approach has unique sensitivity to models with non-\pT-balanced resonances, as well as to models that predominantly couple to gluons, as discussed in Section~\ref{sec:intro}.

%
%
%

% End of text imported from the .//sections/Results.tex input file

%

%
%
%

%
\section{Conclusion}
\label{sec:conclusion}

% The next lines are included from the .//sections/Conclusion.tex input file
This paper presents a search for low-mass dijet resonances in a novel dataset comprising pile-up collisions, obtained using single-lepton triggers and corresponding to 1.30~pb$^{-1}$ of data recorded by the ATLAS detector at the LHC.
The data do not show significant excesses over the Standard Model background prediction.
Two models are used in order to extract 95\% $\mathrm{CL}$ limits on new physics phenomena.
For simplified dark matter $Z'$ models, masses between 135 and 185~GeV are excluded for quark couplings $g_q=0.6$, and the lowest excluded coupling is $g_q=0.28$ at a $Z'$ mass of around 150~GeV.

For signal scenarios that produce Gaussian-like bumps in the dijet mass spectrum, upper limits on $\sigma\times A\times \mathrm{B}$ range between 1.5~nb and 15~nb, depending on the mass, for resonances with a 10\% relative width, and increasing with the width.
Even though Standard Model resonances produced by hadronic decays of $W$ and $Z$ bosons are within the mass range explored, they are not observed. Upper limits at 95\% CL are derived on their yield, and the achieved sensitivity lies near the threshold required for their detection.
This analysis complements the ATLAS low-mass search programme by providing an inclusive search with full object reconstruction and without assumptions on the presence of initial-state radiation.

% End of text imported from the .//sections/Conclusion.tex input file

%
%
%
%
%
\section*{Acknowledgements}
%

%

% The next lines are included from the .//acknowledgements/Acknowledgements.tex input file
%
%

%
%

We thank CERN for the very successful operation of the LHC and its injectors, as well as the support staff at
CERN and at our institutions worldwide without whom ATLAS could not be operated efficiently.

The crucial computing support from all WLCG partners is acknowledged gratefully, in particular from CERN, the ATLAS Tier-1 facilities at TRIUMF/SFU (Canada), NDGF (Denmark, Norway, Sweden), CC-IN2P3 (France), KIT/GridKA (Germany), INFN-CNAF (Italy), NL-T1 (Netherlands), PIC (Spain), RAL (UK) and BNL (USA), the Tier-2 facilities worldwide and large non-WLCG resource providers. Major contributors of computing resources are listed in Ref.~\cite{ATL-SOFT-PUB-2026-001}.

We gratefully acknowledge the support of ANPCyT, Argentina; YerPhI, Armenia; ARC, Australia; BMWFW and FWF, Austria; ANAS, Azerbaijan; CNPq and FAPESP, Brazil; NSERC, NRC and CFI, Canada; CERN; ANID, Chile; CAS, MOST and NSFC, China; Minciencias, Colombia; MEYS CR, Czech Republic; DNRF and DNSRC, Denmark; IN2P3-CNRS and CEA-DRF/IRFU, France; SRNSFG, Georgia; BMFTR, HGF and MPG, Germany; GSRI, Greece; RGC and Hong Kong SAR, China; ICHEP and Academy of Sciences and Humanities, Israel; INFN, Italy; MEXT and JSPS, Japan; CNRST, Morocco; NWO, Netherlands; RCN, Norway; MNiSW, Poland; FCT, Portugal; MNE/IFA, Romania; MSTDI, Serbia; MSSR, Slovakia; ARIS and MVZI, Slovenia; DSI/NRF, South Africa; MICIU/AEI, Spain; SRC and Wallenberg Foundation, Sweden; SERI, SNSF and Cantons of Bern and Geneva, Switzerland; NSTC, Taipei; TENMAK, T\"urkiye; STFC/UKRI, United Kingdom; DOE and NSF, United States of America.

Individual groups and members have received support from BCKDF, CANARIE, CRC and DRAC, Canada; CERN-CZ, FORTE and PRIMUS, Czech Republic; COST, ERC, ERDF, Horizon 2020 and Marie Sk{\l}odowska-Curie Actions, European Union; Investissements d'Avenir Labex, Investissements d'Avenir Idex and ANR, France; DFG and AvH Foundation, Germany; Herakleitos, Thales and Aristeia programmes co-financed by EU-ESF and the Greek NSRF, Greece; BSF-NSF and MINERVA, Israel; NCN and NAWA, Poland; La Caixa Banking Foundation, CERCA and AGAUR programs from Generalitat de Catalunya and PROMETEO and GenT Programmes Generalitat Valenciana, Spain; G\"{o}ran Gustafssons Stiftelse, Sweden; The Royal Society and Leverhulme Trust, United Kingdom; Eric and Wendy Schmidt Fund for Strategic Innovation, United States of America.

In addition, individual members wish to acknowledge support from Chile: Agencia Nacional de Investigaci\'on y Desarrollo (ANID FONDECYT reg. 1230987, FONDECYT 1230812, FONDECYT 1240864, Fondecyt 3240661, Fondecyt Regular 1240721); China: Chinese Ministry of Science and Technology (MOST-2023YFA1605700, MOST-2023YFA1609300), National Natural Science Foundation of China (NSFC 12275265, NSFC-W2543005); Czech Republic: Czech Science Foundation (GACR - 24-11373S), Ministry of Education Youth and Sports (ERC-CZ-LL2327, FORTE CZ.02.01.01/00/22\_008/0004632); EU: H2020 European Research Council (ERC - 101002463); European Union: European Research Council (BARD No. 101116429, ERC - 101219398, ERC 101089007), European Regional Development Fund (HE COFUND GA No.101081355, ERDF), Marie Sklodowska-Curie Actions (GAP-101168829); France: Agence Nationale de la Recherche (ANR-21-CE31-0013, ANR-22-EDIR-0002, ANR-24-CE31-0504-01); Germany: Deutsche Forschungsgemeinschaft (DFG - 469666862); China: Research Grants Council (GRF); Italy: Istituto Nazionale di Fisica Nucleare (LHC-MIUR - 28003/2025), Ministero dell'Università e della Ricerca (NextGenEU I53D23000820006 M4C2.1.1, SOE2024\_0000023); Japan: Japan Society for the Promotion of Science (JSPS KAKENHI  JP25H0063, JSPS KAKENHI JP22H04944, JSPS KAKENHI JP24K23939, JSPS KAKENHI JP24KK0251, JSPS KAKENHI JP25H00650, JSPS KAKENHI JP25H01291, JSPS KAKENHI JP25K01011, JSPS KAKENHI JP25K01023, JSPS KAKENHI JP25KK0047); Poland: Polish National Science Centre (NCN 2021/42/E/ST2/00350, NCN OPUS 2023/51/B/ST2/02507, NCN OPUS nr 2022/47/B/ST2/03059, NCN UMO-2019/34/E/ST2/00393, UMO-2022/47/O/ST2/00148, UMO-2023/49/B/ST2/04085, UMO-2023/51/B/ST2/00920, UMO-2024/53/N/ST2/00869); Spain: Agència de Gestió d'Ajuts Universitaris i de Recerca. (AGAUR - 2023 BP 00141), Ministry of Science and Innovation (RYC2019-028510-I, RYC2020-030254-I, RYC2021-031273-I, RYC2022-038164-I), Ministerio de Ciencia, Innovación y Universidades/Agencia Estatal de Investigaci\'on (EU NextGenerationEU (PRTR-C17.I1), PID2022-142604OB-C22); Sweden: Carl Trygger Foundation (Carl Trygger Foundation CTS 22:2312), Swedish Research Council (Swedish Research Council 2023-04654, VR 2021-03651, VR 2022-03845, VR 2022-04683, VR 2023-03403, VR 2024-05451, VR 2025-05940), Knut and Alice Wallenberg Foundation (KAW 2023.0366); United Kingdom: The Binks Trust, Royal Society (NIF-R1-231091); United States of America: U.S. Department of Energy (ECA DE-AC02-76SF00515), John Templeton Foundation (John Templeton Foundation 63206), Neubauer Family Foundation.

%
%

% End of text imported from the .//acknowledgements/Acknowledgements.tex input file

%
%
%
%
%
%

%
%

%
%
%
\printbibliography
\clearpage
% ATLAS Collaboration author list
% Reference date of EXOT-2022-32 is 2026-04-09
% Author list last updated on date 02-JUN-26
% Data extracted on 02-Jun-2026 for paper reference EXOT-2022-32
% at 8:14am
 
\begin{flushleft}
\hypersetup{urlcolor=black}
{\Large The ATLAS Collaboration}

\bigskip

\AtlasOrcid[0000-0002-6665-4934]{G.~Aad}$^\textrm{\scriptsize 102}$,
\AtlasOrcid[0000-0001-7616-1554]{E.~Aakvaag}$^\textrm{\scriptsize 17}$,
\AtlasOrcid[0000-0002-5888-2734]{B.~Abbott}$^\textrm{\scriptsize 121}$,
\AtlasOrcid[0000-0002-0287-5869]{S.~Abdelhameed}$^\textrm{\scriptsize 83b}$,
\AtlasOrcid[0000-0002-1002-1652]{K.~Abeling}$^\textrm{\scriptsize 54}$,
\AtlasOrcid[0000-0001-5763-2760]{N.J.~Abicht}$^\textrm{\scriptsize 48}$,
\AtlasOrcid[0000-0002-8496-9294]{S.H.~Abidi}$^\textrm{\scriptsize 30}$,
\AtlasOrcid[0009-0003-6578-220X]{M.~Aboelela}$^\textrm{\scriptsize 44}$,
\AtlasOrcid[0000-0002-9987-2292]{A.~Aboulhorma}$^\textrm{\scriptsize 36e}$,
\AtlasOrcid[0000-0001-5329-6640]{H.~Abramowicz}$^\textrm{\scriptsize 154}$
% ...
\bigskip
\\

% ...
$^{17}$Department for Physics and Technology, University of Bergen, Bergen; Norway.\\
% ...
$^{30}$Physics Department, Brookhaven National Laboratory, Upton NY; United States of America.\\
% ...
$^{36}$$^{(a)}$Facult\'e des Sciences Ain Chock, Universit\'e Hassan II de Casablanca;$^{(b)}$Facult\'{e} des Sciences, Universit\'{e} Ibn-Tofail, K\'{e}nitra;$^{(c)}$Facult\'e des Sciences Semlalia, Universit\'e Cadi Ayyad, LPHEA-Marrakech;$^{(d)}$LPMR, Facult\'e des Sciences, Universit\'e Mohamed Premier, Oujda;$^{(e)}$Facult\'e des sciences, Universit\'e Mohammed V, Rabat;$^{(f)}$Institute of Applied Physics, Mohammed VI Polytechnic University, Ben Guerir; Morocco.\\
% ...
$^{44}$Physics Department, Southern Methodist University, Dallas TX; United States of America.\\
% ...
$^{48}$Fakult\"{a}t Physik, Technische Universit{\"a}t Dortmund, Dortmund; Germany.\\
% ...
$^{54}$II. Physikalisches Institut, Georg-August-Universit\"{a}t G\"ottingen, G\"ottingen; Germany.\\
% ...
$^{83}$$^{(a)}$Khalifa University of Science and Technology, Abu Dhabi;$^{(b)}$New York University Abu Dhabi, Abu Dhabi;$^{(c)}$United Arab Emirates University, Al Ain;$^{(d)}$University of Sharjah, Sharjah; United Arab Emirates.\\
% ...
$^{102}$CPPM, Aix-Marseille Universit\'e, CNRS/IN2P3, Marseille; France.\\
% ...
$^{121}$Homer L. Dodge Department of Physics and Astronomy, University of Oklahoma, Norman OK; United States of America.\\
% ...
$^{154}$Raymond and Beverly Sackler School of Physics and Astronomy, Tel Aviv University, Tel Aviv; Israel.\\
% ...
$^{a}$ Also at Affiliated with an institute formerly covered by a cooperation agreement with CERN.\\
$^{b}$ Also at An-Najah National University, Nablus; Palestine.\\
$^{c}$ Also at Borough of Manhattan Community College, City University of New York, New York NY; United States of America.\\
$^{d}$ Also at Center for Interdisciplinary Research and Innovation (CIRI-AUTH), Thessaloniki; Greece.\\
$^{e}$ Also at Centre of Physics of the Universities of Minho and Porto (CF-UM-UP); Portugal.\\
$^{f}$ Also at CERN, Geneva; Switzerland.\\
% ...

\end{flushleft}

% Created with Glance <Atlas.Glance@cern.ch>

%

%
%

%

%
%
%
%

\end{document}